\documentclass{article}

\usepackage{arxiv}

\usepackage[utf8]{inputenc} 
\usepackage[T1]{fontenc}    
\usepackage{hyperref}       
\usepackage{url}            
\usepackage{booktabs}       
\usepackage{amsfonts}       
\usepackage{nicefrac}       
\usepackage{microtype}      
\usepackage{lipsum}
\usepackage{graphicx}
\usepackage[dvipsnames]{xcolor}

\title{Hypersyn: A Peer-to-Peer System for Mutual Credit}
\author{Lum Ramabaja \\ \href{mailto:lum@hypersyn.org}{lum@hypersyn.org} \\ Draft Version 0.0.1 }
\date{March 2022}

\begin{document}

\maketitle

\begin{abstract}
    The Hypersyn protocol is a new type of permissionless and peer-to-peer payment network that is based on the concept of mutual credit and mutual arbitrage. Unlike blockchain-based systems, Hypersyn does not rely on any consensus algorithm. It does not require a distributed ledger to store the history of events nor a set of validators. Hypersyn does not have a system-imposed hard-cap on the number of transactions per second that it can perform, and can therefore easily scale up or down depending on network usage. Unlike in other payment systems, money in Hypersyn does not get transferred from person $A$ to person $B$ in the conventional sense. Instead of transferring a token between each other, peers in Hypersyn change the exchange value of their credit (i.e. their purchasing power) within the network. Just as in centrally-issued fiat systems, money in Hypersyn is treated as freely tradable debt, which inherently requires trust. But unlike centrally-issued fiat systems, money issuance in Hypersyn is not controlled by an authority, but is instead created on the spot as mutual credit. In blockchain-based systems and even in centrally-issued fiat systems, money is treated as a scarce commodity. In the Hypersyn protocol on the other hand, money supply within the system is elastic in nature. Because of these fundamental differences in assumptions, the Hypersyn protocol does not aim to compete with, or substitute blockchain-based systems. Instead, Hypersyn should be viewed as a tool that aims to offer a qualitative change in the way we exchange. It has the potential to increase the autonomy and self-organization that people can have, by enabling people to become both the creditors and debtors of their own "money" through mutual credit.
\end{abstract}



\section{Introduction}
The capacity of a social system is intertwined with the limitations of its monetary system. By allowing banks to have absolute control of credit, we have inevitably limited the level of autonomy and self-organization that people can have. To break this power asymmetry, we need new tools and systems that can democratize credit. The Hypersyn protocol intends to achieve this by enabling people to become both the creditors and debtors of their own "money" through mutual credit. The first part of the introduction (subsection \ref{ss:self_issued_credit}) will briefly discuss about the author's understanding of money and different forms of money throughout history. This subsection introduces historical examples to help understand money as freely tradable debt. The second part of the introduction (subsection \ref{ss: sparse_merkle_tree} will briefly introduce the concept of Merkle trees as well as sparse Merkle trees. sparse Merkle trees represent the data structure of choice used for Hypersyn's state model. The second part of the paper focuses on the Hypersyn protocol overall. Subsection \ref{ss: overview} briefly describes the main purpose of Hypersyn. Subsection \ref{ss: exchange mechanism} explores in more detail how payments in Hypersyn work, as well as explains Hypersyn's relation to existing automated market maker models. Subsection \ref{ss: mutual arbitrage} introduces the concept of \textit{mutual arbitrage}, an agent-centric arbitrage mechanism used to update the exchange value of credits within the Hypersyn network. Subsection \ref{ss: state model} describes Hypersyn's state model. Here we explain how Hypersyn makes use of sparse Merkle trees and pruned sparse Merkle trees to manage and store the exchange value of peer credit. Subsection \ref{ss: edge formation} briefly describes how nodes in the Hypersyn network can form and remove edge connections with one-another. The Hypersyn files subsection (Subsection \ref{ss: network}) introduces how a torrent-file-like approach is used for first time Hypersyn contacts. In section \ref{s: discussion} we both cover Hypersyn's implications for the existing monetary system, as well as its implications for blockchain-based systems.

\subsection{Money \& Mutual Credit}
\label{ss:self_issued_credit}

In medieval Europe, "split tally sticks" were used by local businesses as temper proof recording devices to create self-issued credit (or government-issued credit) in exchange for commodities and services \cite{goetzmann_origins_2005}. The split tally sticks were made out of wood and had no intrinsic value. They simply encoded the amount of debt an entity owed. The information got encoded by carving horizontal lines onto the stick and then splitting the stick down its  length in halve. One half (known as the "foil") would be kept by the debtor. The other half (known as the "stock") was kept by the creditor. To prove that the amount was not tempered with, all one had to do, was to align the two halves together to check the engravings. If entity $Y$ had a tally stock containing the information "$X$ owes 100 gold coins to $Y$", as long as $X$ was considered trustworthy, the tally stock itself was worth 100 gold coins and could be traded freely with other entities. \textit{The tally stock, that is the debt, became money}. This is why we can refer to money as debt that is freely tradable.  

Other regions of the world had also periods when self-issued credit was used. In China for example, between the 1870s until the 1940, businesses would frequently issue their own money (also known as bamboo tally, or bamboo money), as emergency money. The money, which essentially was self-issued credit, did not only insulate local businesses from the economic decline of a weakened government, but enabled the economic self organization of local communities. Local trade was still possible, even if the state's currency failed. Historically however, self-issued credit was rarely issued as tangible money. Tally sticks are rather the exception than the norm. Instead of minting money in the form of coins or tallies, many societies simply recorded transactional events as credit in a ledger directly. 

We know for example that many societies in the region of Andean South America used "quipus" for bookkeeping and accounting \cite{benson_quipu_1975}. The quipu was a mnemonic device (i.e. an information storage and retrieval device) composed of several knotted strings. Each sequence of knot configurations encoded a different symbolic meaning. It is believed that the quipu may have carried far more types of information than only bookkeeping \cite{benson_quipu_1975}. Individuals in European villages often did something similar as the South American cultures: Instead of actually minting a coin, or a tally when transacting with one-another, they recorded transactional events as credit in a village ledger directly (a book that kept track on how much anyone owes anyone else). Note that for any practical purposes, the manifestation of credit was irrelevant. It did not matter if the credit was represented as a tally, as a coin, as a string of knots, or as a record in a village book. What mattered was that the recorded debt was equivalent to having "money", because of \textit{credit clearing}.

Credit clearing is the practice where the credits of multiple parties are cancelled out with each other at regular time intervals. Or in other words, it is the process where payments get cancelled by other payments that come from the "opposite direction". In the 1600s for example,  credit clearance was used for a long time between the banks of London in England. Each bank interacted with other banks without using money, they simply kept a tally of balances for each interaction. At the end of each day, the banks would send their debt balances to the clearing house, where the debts would be settled and only a single payment to or from each bank had to be made. A form of credit clearance also happens within the payment channels of the Bitcoin lightning network \cite{poon_bitcoin_2016}. Two lightning nodes can transact off-blockchain with one-another as often as they want and only make an actual Bitcoin payment after sending the proofs to the Bitcoin "clearing house". This is similar to how banks today settle within the shared ledger of the central bank.

Credit clearing systems are often also referred to as \textit{mutual credit} systems. Mutual credit is exactly how villagers throughout history payed each other for commodities and services. Instead of using state-issued fiat for their payments, they used mutual credit by keeping track of each-others credits through tallies, strings, books, or other technologies. In mutual credit systems, the "money" to mediate a transaction is created on the spot as a corresponding credit and debit in the balances of the two parties. This particular design choice has several interesting implications:
\begin{enumerate}
    \item Systems that use mutual credit are more resilient towards external shocks, than systems with centrally-issued fiat. In systems with centrally-issued fiat, when economic times are tough, everyone suffers together. When a collapse occurs, centrally-issued fiat systems experience a top-down collapse. Local supply chains break down even if the root causes of the crisis are at the top. Mutual credit systems are by design decentralized and cannot experience a top-down collapse. Local supply chains can remain functional, even if larger social structures fail. Collapse in mutual credit systems is local in nature, instead of top-down.
    \item Centrally-issued fiat is a scarce commodity, which incentivizes players to hoard it. This in turn leads to gatekeeping effects, creating a feedback loop. By controlling and restricting the behaviors and access of people, the ruling class can hoard even more wealth.
    In mutual credit systems, as every credit is matched by an equal and opposite debt, credit cannot be hoarded. Mutual credit systems furthermore are designed to have a perfectly elastic supply. It is available whenever a trade is needed. One's purchasing power dynamically increases, or decreases, depending on the amount of debt they have in the system.
    \item In centrally-issued fiat systems, banks (and therefore the state) are both the issuer as well as the collectors of their own debt. Just as capitalists have absolute control over the means of production, banks have absolute control over credit. By enforcing centrally-issued fiat onto a population, the state, influenced by the ruling class (the state is after all an instrument for the ruling class) has the power to perform frictionless surplus value extraction from its working class, as well as perform behavioral control on its working class. By having a monopoly on credit, the ruling class is able to destroy capital, as well as move it around as desired. This power asymmetry does not exist in mutual credit systems. Hierarchies in mutual credit systems are functional in nature (view the "functional hierarchies" subsection from \cite{swann_towards_2018}). If a central-authority misuses its power, its credit within the system loses purchasing power. Capital can also not be moved around or destroyed, because the exchange value of one's credit is not defined by a centrally-issued currency, but is directly defined as a relation between the labor of two (or more) parties. Therefore mutual credit systems are a direct representation of real-world production relations, rather than abstract claims on collective future productivity in the form of collective debt.
\end{enumerate}

Unlike what many textbooks claim, it was not centrally-issued fiat that historically solved the coincidence of wants problem \cite{szabo_shelling_2002}, but mutual credit. Coincidence of wants is an economic phenomenon in which two (or more) parties perform a exchange mechanism of assets between themselves, without the use of any money (also known as bartering). Because of the improbability of wants however (one party might simply not have a product, or service that the other party wants, or needs), bartering between parties often appears infeasible. Credit as an innovation solved for the coincidence of wants problem, by adding a temporal dimension to the bartering process. Instead of exchanging one product for another, individuals were able to exchange one product for credit, which could then be cleared out at a given future time. Credit is therefore a time-delayed multi-agent bartering system, with a mutually agreed upon denominator and redemption mechanism. Mutual credit, not centrally-issued fiat, was for the longest time the default monetary system in the world.

There is however a catch in mutual credit systems. For a party $X$ to be able to spend another party's ($Y$) debt with party $Z$, $Z$ has to believe that $Y$ is trustworthy and that their debt will eventually be repaid. Money, or debt, therefore inherently requires \textit{trust}. And this is where things become tricky. Game theoretically speaking, trust can only exist between players that play "repeated games" (i.e. that interact more than once with one another). Trustworthiness therefore decays the larger the network size of players becomes. Due to what is most likely a biological constraint (there is an upper cap of long lasting connections that a player can form), the more players get added to the social graph, the more likely one-time games become between players. In other words, the more one-time games happen, the less likely it gets for $Z$ to trust that another player's credit can get eventually cleared out. Trust erodes with network size. Mutual credit, as it appeared historically, was fundamentally not scalable. 

As a technology, mutual credit worked in smaller collectives and communities, but once the network size of a society surpassed a certain threshold, individuals were incentivized to adopt centrally-issued money. By paying with money that was centrally-issued, players (especially those players that were part of the same state) were able to trade the state's money with other players, without having to trust the other player. This greatly simplified things for individuals, but also gave the ruling class unprecedented power. 

One thing is important to be kept in mind: Modes of production influence the way we explore the technological search space. For example, if our goal is to optimize for surplus value alone, then the technologies that we create will themselves have the criteria of surplus value maximisation as part of their design. But the opposite relation is also true - inventing new technologies and tools can have a direct influence on how societies organize and interact. This is the main aim of the the Hypersyn protocol: It is a tool that aims to offer a qualitative change in the way we exchange. It is a tool to enable mutual credit systems to scale beyond small communities.

\subsection{Sparse Merkle Tree}
\label{ss: sparse_merkle_tree}

\begin{figure}[h]
\centering\includegraphics[width=0.5\linewidth]{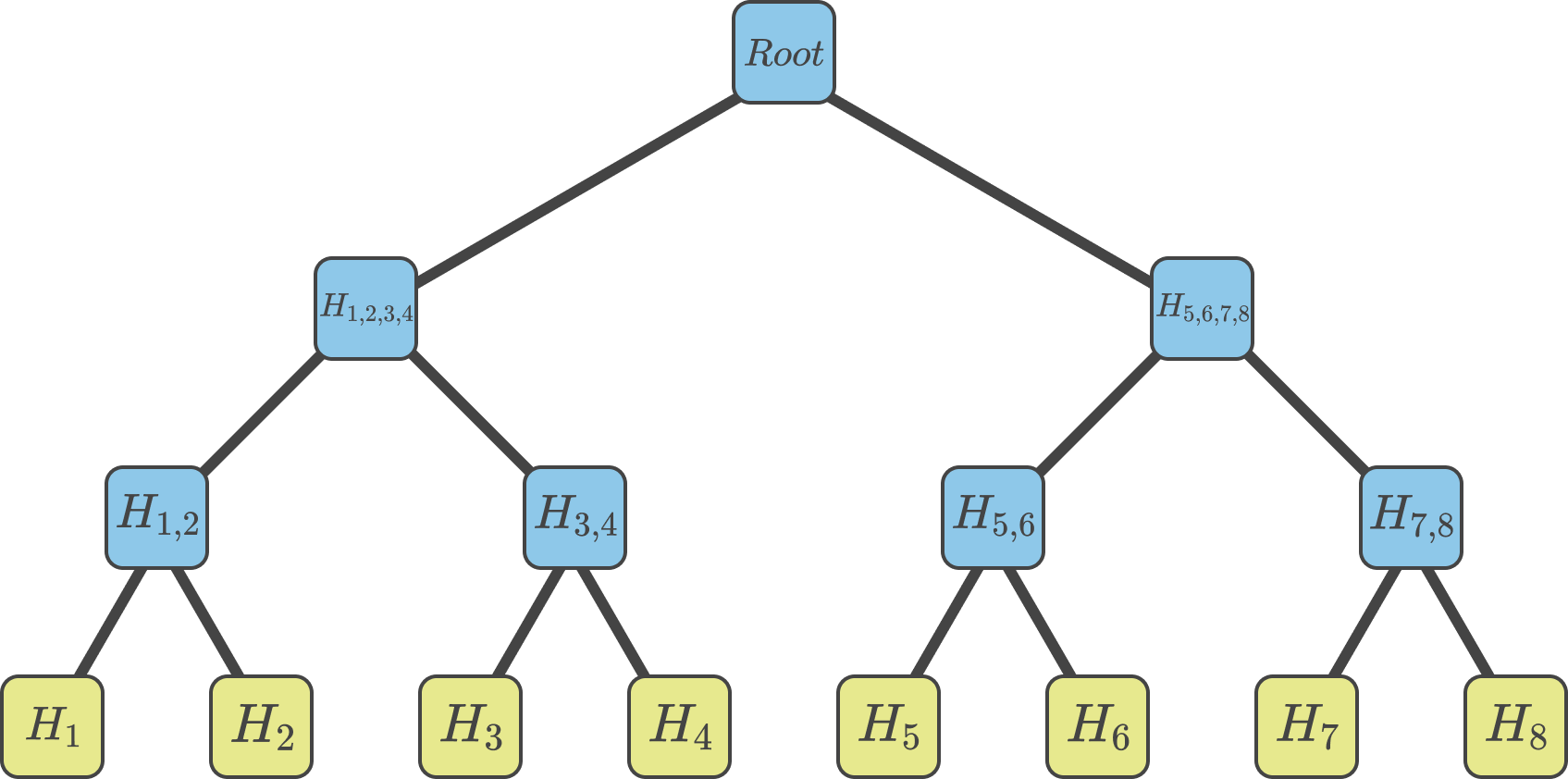}
\caption{A depiction of a Merkle Tree. Leaf nodes are represented in yellow. Non-leaf nodes are represented in blue.}
\label{fig:Merkle_tree}
\end{figure}

Hypersyn's key component for its state model is the sparse Merkle tree. This subsection briefly introduces the concept of a Merkle tree and a Sparse Merkle tree. How the sparse Merkle tree is used in Hypersyn is described in subsection \ref{ss: state model}. A Merkle tree is a binary tree data structure used to securely verify the presence of values in a list, without having to provide every value of the list to another party. To construct a Merkle tree, every value is hashed via a cryptographic hash function, the transformed values are also known as \textit{leaf nodes}. The leaf nodes are then hashed together to form parent nodes, also known as \textit{non-leaf nodes}. This step is repeated until the root of the constructed binary tree is reached (see Figure \ref{fig:Merkle_tree}).

\begin{figure}[h]
\centering\includegraphics[width=0.5\linewidth]{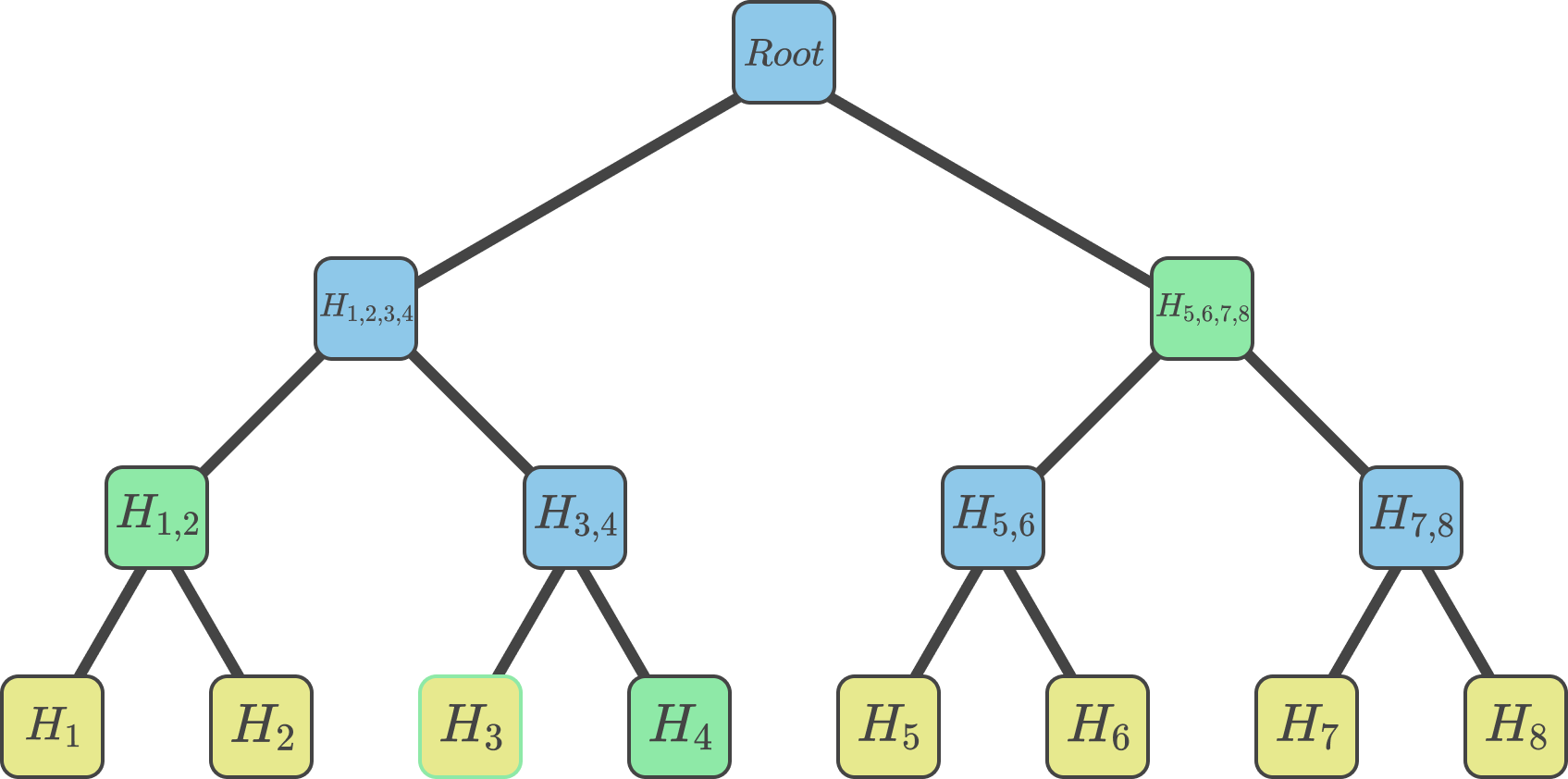}
\caption{A depiction of a Merkle proof. To prove that $H_3$ was present in the initial value list, one has to hash it with $H_4$, then with $H_{1,2}$, and finally with $H_{5,6,7,8}$ (shown in green) to recreate the hash of the Merkle root.}
\label{fig:Merkle_proof}
\end{figure}

To verify that a value is present in the Merkle tree, a series of hashes (also known as the \textit{Merkle proof}) is provided. By sequentially hashing the leaf node hash with the provided Merkle proof, the Merkle root of the original Merkle tree can be recreated (see Figure \ref{fig:Merkle_proof}). Note that recipients of a Merkle proof must already have a local copy of the Merkle root for a proof to take place. A node that receives a Merkle proof therefore, is able to verify if a value was part of the list of values that generated the Merkle root, by comparing their locally stored Merkle root, with the final hash generated by the provided Merkle proof. If the two hashes are equivalent, then the recipient of the Merkle proof knows that the provided value was indeed one of the leaf nodes in the original Merkle tree. This is also known as a \textit{presence proof}. Instead of having to provide the whole list of $M$ values to another party to verify the presence of one specific value, Merkle trees allow to prove the presence of a value by providing only $\log{M}$ hashes.

While Merkle trees can provide presence proofs and perform value insertions with $\log{M}$ steps, they do not allow for efficient \textit{absence proofs}, value updates, and value deletions. An absence proof is when a Merkle proof can show the absence of a value in a tree. In regular Merkle trees, absence proofs require leaf node sorting, while value updates and deletions require the re-computation of the whole tree, all of which are computationally expensive. The sparse Merkle tree (SMT) solves these issues by slightly redefining how a tree is constructed. In the SMT, one can both provide presence and absence proofs in $\log{M}$ steps on average, as well as perform insertions, deletions, and updates in $\log{M}$ steps on average.

\begin{figure}[h]
\centering\includegraphics[width=0.8\linewidth]{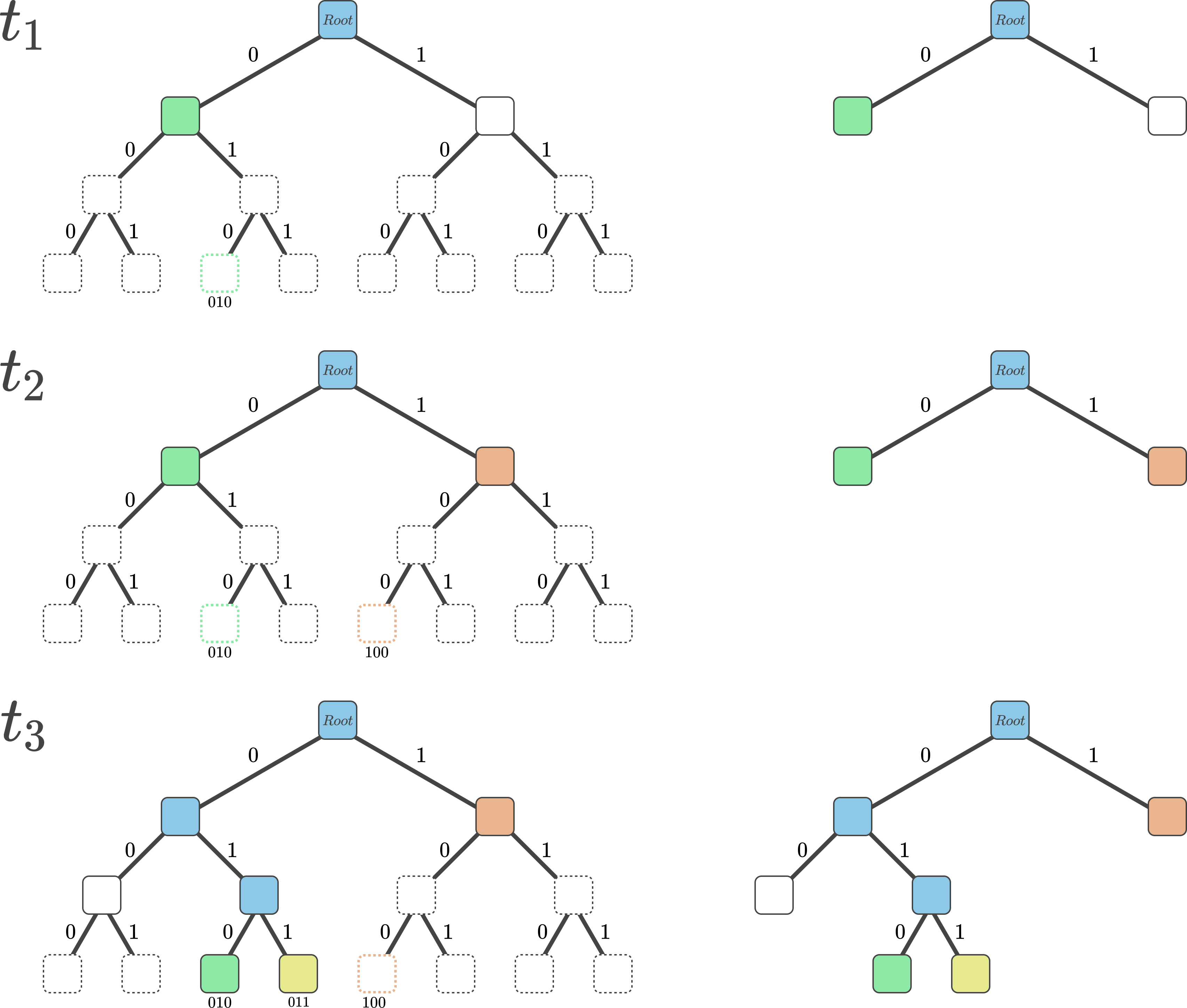}
\caption{A depiction of a sparse Merkle tree. At every time step $t$ a value is being inserted. Blue nodes represent non-leaf nodes. White nodes with full lines represent empty nodes. White nodes with dashed lines are not part of the SMT and serve to depict where value nodes in the tree would get mapped. The green ($010$), orange ($100$), and yellow ($010$) nodes represent the values, i.e. leaf nodes of the SMT. }
\label{fig:SMT_insertion}
\end{figure}

Instead of constructing a Merkle tree bottom-up from a fixed set of values, SMTs start with an empty tree and sequentially add values to it. This is done by introducing the concept of \textit{empty nodes} (a constant node with a fixed hash value), as well as by following a simple heuristic: 
\begin{enumerate}
 \item To add a new value to the tree, begin from the tree root. If there is no value yet in the tree, let the tree root be the empty node.
 \item Let the path from a parent node to the left child node represent a zero, and the path to the right child node represent a one (see Figure \ref{fig:SMT_insertion}). Every value has a bit sequence (their cryptographic hash) that can be mapped onto the tree by sequentially choosing either the left side, or right side of a parent node, depending on the bits of the value's hash. Go down the value's path in the tree until a leaf node, or an empty node is reached.
 \item If an empty node is reached, substitute it with the new value. Taking Figure \ref{fig:SMT_insertion}) as an example, inserting the orange value with ID $100$ (at time step $t_2$) will result in the substitution of an empty node with the actual value.
 \item If a leaf node is reached, create two child nodes and remap the values (by looking at the hashes of the values). If the values get mapped to the same child node, repeat the process and insert an empty node at the opposite child node. Taking Figure \ref{fig:SMT_insertion}) as an example, inserting the yellow value with ID $011$ (at time step $t_3$) will result in the repositioning of the value with ID $010$, as well as result in the creation of an empty node on the left side of the none-leaf node that was previously (in time step $t_2$) the $010$ leaf node.
\end{enumerate}

\begin{figure}[h]
\centering\includegraphics[width=0.8\linewidth]{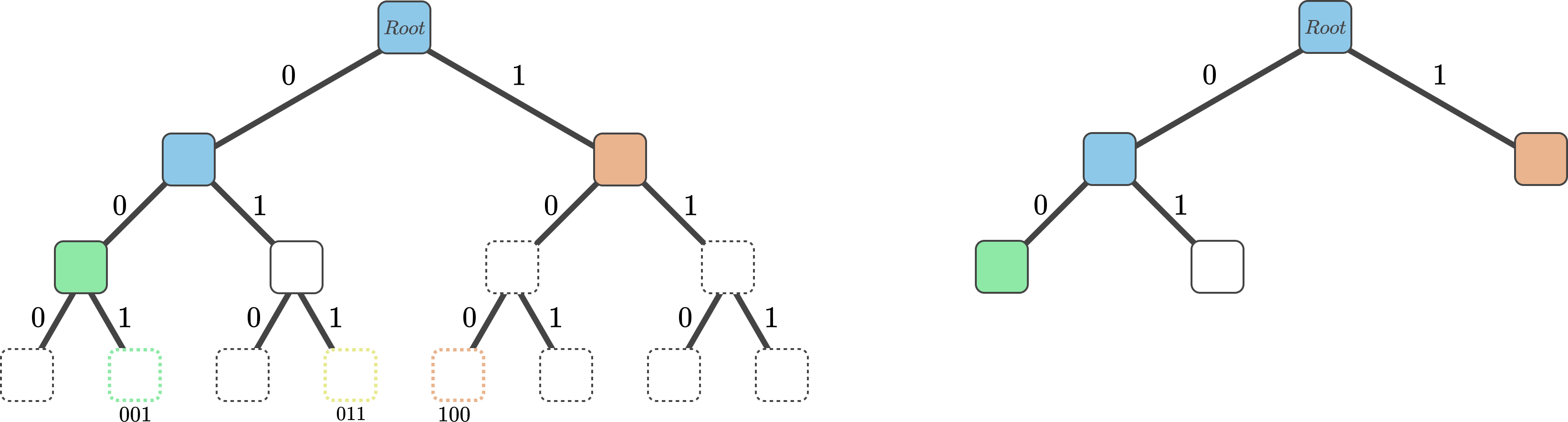}
\caption{A depiction of a sparse Merkle tree. the value with ID $011$ was never inserted into the SMT. One can show this by generating an absence proof.}
\label{fig:SMT_absence}
\end{figure}

Value deletions in SMTs work the exact same way as insertions, but in reverse. When removing a leaf node we move any sibling node up, until a parent node with another leaf node, or empty node is reached. Proving the presence of a value in SMTs is done the same way as in regular Merkle trees (see Figure \ref{fig:Merkle_proof} as reference), but surprisingly enough, so are absence proofs. To prove that a value was not part of a sparse Merkle tree, all one has to do is provide a regular Merkle proof, where the original leaf value of the proof represents the empty node value (a constant hash value). Taking Figure \ref{fig:SMT_absence} as an example, to prove that the value with ID $001$ (depicted with a yellow dashed line) is not part of the SMT, one simply has to provide the empty node, as well as the green leaf node and the orange leaf node to a recipient. By hashing the empty node value with the green node value and the orange node value, one recreates the original Merkle root and can therefore prove that any value that starts with $01$ cannot have been part of the tree.

In agent-centric protocol designs, nodes usually sign their Merkle roots with their private keys before sending it to peers. This is done to prevent malicious peers from forging another node's Merkle roots. When the volume of requests on a node-basis is high however, that is, when a node has to update, insert, and/or delete many of its SMT values in a small amount of time, signing the Merkle root after every state change becomes costly. To overcome this scalability issue, nodes in Hypersyn can batch-sign a number of transactions (i.e. SMT state changes) by signing a single Merkle root and a counter, as long as the state changes have a partial order with each other (see section \ref{ss: state model}).

\section{Hypersyn}
\label{hypersyn}

\subsection{Overview}
\label{ss: overview}

In conventional money systems, such as centrally-issued fiat currencies or blockchain-based cryptocurrencies, money is often treated as an "object" that moves from one party to another when transacting. This is also why blockchain-based systems require a consensus algorithm - the nodes in the network have to arrive independently to an absolute (or in the case of decentralized acyclic graph architectures, eventual) order of events. To achieve that, the distributed ledger needs to have one single state. In other words, the system has to ensure that no \textit{double spending} can occur. If party $A$ sends an "object" to party $B$, the system has to ensure that the same object cannot be sent at the same time to party $C$. The role of consensus algorithms in blockchain-based systems therefore is to ensure the consistency of the system's state across nodes. This approach does come however with its drawbacks. Current blockchain-based systems like Bitcoin \cite{nakamoto_bitcoin_2008}, or Ethereum \cite{wood_ethereum_2013} as of the time of this writing can only process around 5 to 15 transactions per second. To increase the number of events (i.e. transactions) within the system, the system can either decide to become more centralized (by increasing block size, or by having fewer validators with more resources, etc.), or by sharding transaction validation (which lowers the security of the network). 

In the case of the Hypersyn network, there is no system-imposed hard-cap on the number of transactions per second that can be processed. Because of the CRDT-like \cite{shapiro_conict-free_2011} nature of sparse Merkle roots, the Hypersyn protocol does not require a consensus algorithm. It substitutes consensus for state. By redefining what money is, we manage to dissolve the double spending problem. By eliminating the double spending problem, we can dissolve the need for consensus and the need to enforce a system-wide absolute order of events. By not requiring validators to check every transaction in the system, we manage to dissolve the need for a distributed ledger. Instead of trying to solve for the problems that blockchains have, Hypersyn tries to redefine them. As Stafford Beer once said: "It is better to dissolve problems than to solve them" \cite{beer_designing_1994}. 

In Hypersyn for example, there is no finite amount of objects (i.e. coins) that flow through the system. Or put differently, "Money" in Hypersyn is not a scarce commodity. Instead, nodes in Hypersyn make use of mutual credit, i.e. the system has an elastic supply of "money". Each credit in the Hypersyn system does not even have the same value. The exchange value of one's credit is determined through the credit reserve ratios within the connection of two peers (see section \ref{ss: exchange mechanism}. Because of the mutual arbitrage process (see section \ref{ss: mutual arbitrage}), what is really being transferred when transacting in Hypersyn is not money in the conventional sense, but a system-wide change in one's exchange value (or purchasing power). As we will see in the next sections, peer-to-peer mutual credit behaves similarly to the credit clearance concept introduced in Section \ref{ss:self_issued_credit}.

Besides solving many of the problems that plague blockchain-based systems, Hypersyn has also some interesting implications for centrally-issued fiat systems. Peer-to-peer mutual credit enables for the first time ever the \textbf{true} "separation of money and state" \cite{riegel_new_1976}. It enables people to transact in something other than the inflation-prone / deflation-prone centrally-issued money of the state. It enables people to become both the creditors and debtors of their own money through mutual credit.

\subsection{Exchange Mechanism}
\label{ss: exchange mechanism}
This section focuses on providing a general overview of the Hypersyn payment mechanism, the exact details of the peer-to-peer protocol are discussed in the subsequent sections. In Hypersyn, as in most peer-to-peer system, the address of a node $N$ is determined through a public address. Each node can create credit $C$ which it can use as payment with other nodes. The credit used by nodes is not backed by a specific commodity, or a basket of commodities, as proposed in previous mutual credit systems \cite{greco_end_2009}. The exchange value of a node's credit, and therefore its purchasing power, is determined through its connections $E$, also known as edges or relations, with other nodes. Each edge in Hypersyn contains a pair of reserves $R$ containing credits from both nodes which it connects. How edges are created and removed from the network will be discussed in section \ref{ss: edge formation}. For the examples of this section, for simplicity's sake, let's assume a network with already existing edges between a subset of nodes. In Hypersyn, nodes cannot \textit{own} another node's credit in the conventional sense. All a node can do is change the exchange value between their credit and that of their peers by modifying the credit reserves of their edges. The way two nodes determine the exchange value between their credits, is by following an agent-centric constant function market maker (CFMM) design. 

CFMMs, which are a type of automated market maker (AMM) \cite{othman_automated_2012}, were originally designed to create decentralized exchanges (DEXs) for digital assets in the blockchain space. In CFMMs, the price of two digital assets is determined by using a \textit{constant function}. A constant function is a function that preserves an invariance. In the case of the constant product market maker model implemented by Uniswap \cite{zhang_formal_2018} and used by Hypersyn, the trade between two assets has to happen in a way that keeps the product of the two asset reserves unchanged after the trade (see Equation \ref{eq:const_prod1}). In Hypersyn however, instead of using the constant function to trade digital assets in a DEX, we use it to make peer-to-peer credit payments between peers, and to determine the exchange values between different credits. The constant function used in Hypersyn can be defined as:

\begin{equation}
\label{eq:const_prod1}
(R_A + \Delta_A)(R_B - \Delta_B) = k
\end{equation}

where $R_A$ and $R_B$ are $E_{AB}$'s credit reserves containing $C_A$ and $C_B$ respectively, $\Delta_B$ shows the credit amount that needs to be inserted into the edge, and $\Delta_A$ shows the credit amount that gets removed from the edge. The constant $k$ has to remain equal before and after the exchange. Or in other words:

\begin{equation}
\label{eq:const_prod2}
(R_A + \Delta_A)(R_B - \Delta_B) = R_A R_B
\end{equation}

Unlike with centrally-issued fiat, the price of an exchange in Hypersyn is different from the perspective of each node. For the producer $N_B$, each of its products and services has a specific price in terms of $C_B$. When $N_A$ however wants to exchange with $N_B$, it will not pay $N_B$ in terms of $C_B$, it will have to pay in terms of $C_A$.

\begin{figure}[h]
\centering\includegraphics[width=0.4\linewidth]{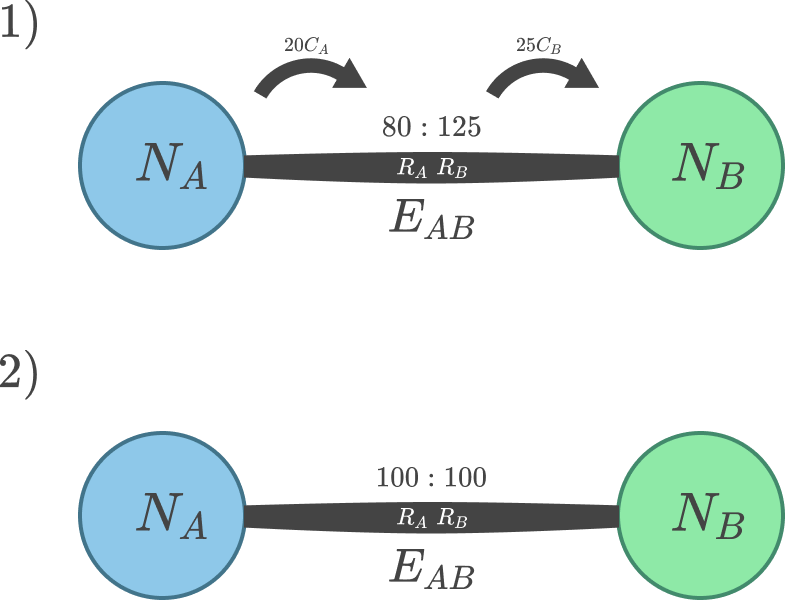}
\caption{A depiction of a Hypersyn payment.}
\label{fig:hypersyn_payment}
\end{figure}

Taking figure \ref{fig:hypersyn_payment} as an example, if we have two nodes $N_A$ and $N_B$ with an edge $E_{AB}$ and if a product or service of $N_B$ costs $25 C_B$, and $E_{AB}$ has an $R_A$ of $80 C_A$ and an $R_B$ of $125 C_B$, $N_A$ would have to create $20 C_A$ and deposit the created amount into $E_{AB}$. $N_B$ would then take out $25 C_B$ from $E_{AB}$, delete the amount from circulation, and provide the requested product or service to $N_A$. How much $C_B$ would be removed from $E_{AB}$ by a given amount of $C_A$, can be found out by calculating:

\begin{equation}
\label{eq:const_prod3}
 \Delta_B= \frac{\Delta_A R_B}{R_A + \Delta_A}
\end{equation}

Where $R_A$ and $R_B$ represent the corresponding credit reserves within $E_{AB}$. To calculate therefore how much credit $N_A$ would have to spend for a given amount of $C_B$, we would use the formula:

\begin{equation}
\label{eq:const_prod4}
 \Delta_A= R_A \frac{\Delta_B}{R_B - \Delta_B}
\end{equation}

From the perspective of $N_A$ therefore, the exchange with $N_B$ has cost $20 C_A$. From the perspective of another node, let's say, $N_C$, that same exchange with $N_B$ might only cost $1 C_C$. In other words, in Hypersyn, there is a different exchange value between the credit of every node, the exchange value of which is determined by the credit reserve ratios of their edges. 

But what if a trade relation between two peers is not bidirectional, as it is often the case in real-world scenarios? One might think that $N_A$ would have to spend ever more credit whenever it makes a payment to $N_B$. After all, the more skewed $E_{AB}$'s credit reserves get, the less exchange value $C_A$ has compared to $C_B$ (ever more $C_A$ have to be spend for a given amount of $C_B$). To prevent unidirectional trade relations from forming ever larger asymmetries within the system, Hypersyn makes use of \textit{mutual arbitrage}.

\subsection{Mutual Arbitrage}
\label{ss: mutual arbitrage}

\begin{figure}[h]
\centering\includegraphics[width=0.75\linewidth]{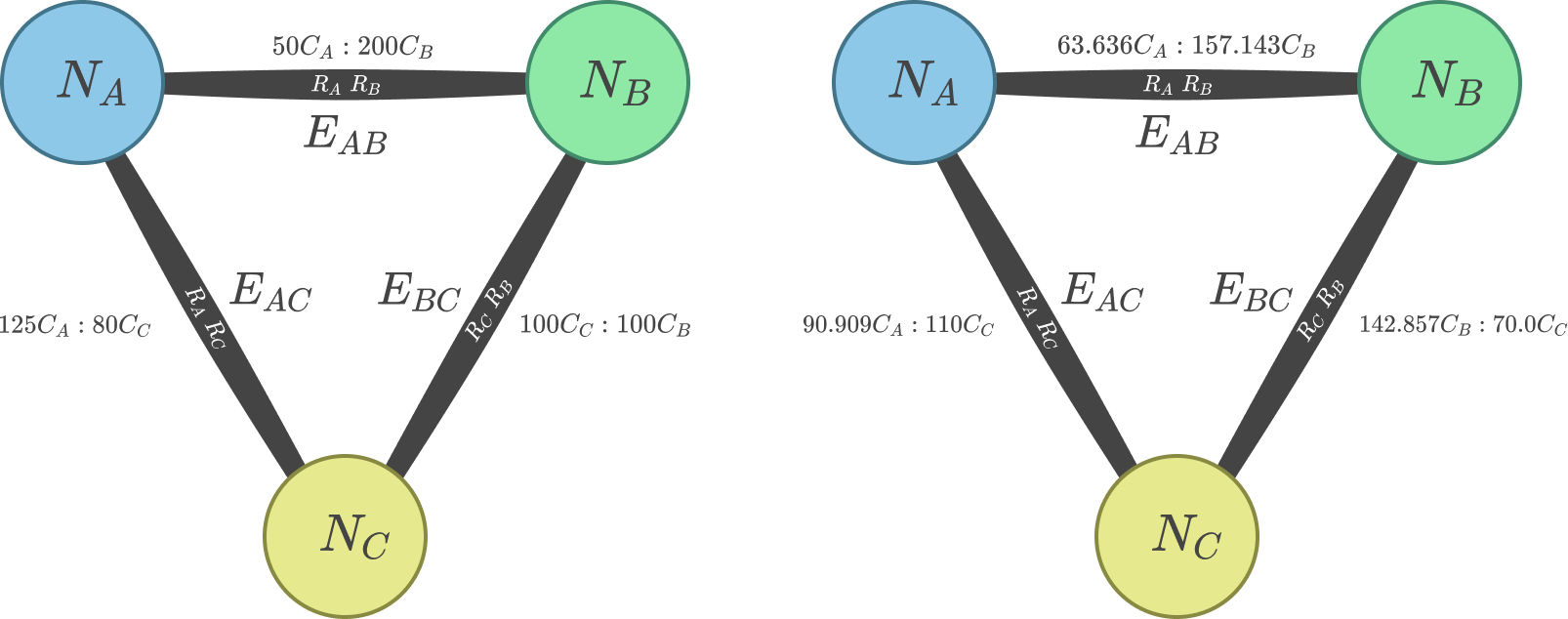}
\caption{A depiction of agent-centric arbitrage in Hypersyn. The left image shows the credit exchange values of three nodes before mutual arbitrage is performed. The right image shows the adjusted credit exchange values of three nodes after mutual arbitrage is performed. Each edge updates its credit reserves after the arbitrage. Note that the reserves of each edge remains invariant, that is $R_i R_j = (R_i + \Delta_i)(R_j - \Delta_j)$
}
\label{fig:mutual_arbitrage}
\end{figure}

Mutual arbitrage in Hypersyn is a form of agent-centric arbitrage that is used to correct and update the exchange values of credits in a peer-to-peer fashion. It behaves similarly to the credit clearance mechanism introduced in section \ref{ss:self_issued_credit}. Mutual arbitrage can happen only between three (or more) nodes that are connected with each-other (see Figure \ref{fig:mutual_arbitrage}). The purpose of mutual arbitrage in Hypersyn is two fold: 
\begin{enumerate}
    \item It keeps the exchange value of a node's credit up to date with its peers. If node $N_B$ consumes significantly more from $N_A$ than it produces for the rest of its peers (e.g. $N_C$), then the value of $C_B$ will decrease between all of $N_B$'s peers after mutual arbitrage. Taking Figure \ref{fig:mutual_arbitrage} as an example, if $N_B$ consumes frequently from $N_A$, their edge $E_{AB}$ will end up having skewed credit ratios. That is, $E_{AB}$ will have significantly more $C_B$ than $C_A$ (left image). After mutual arbitrage between $N_A$, $N_B$, and $N_C$ however (right image), the credit reserves inside $E_{AB}$ become slightly less skewed (i.e. $N_B$'s purchasing power in relation to $N_A$ slightly increased from before). Through periodic mutual arbitrage, the exchange value of each node's credit remains up to date and accurately reflects the consumption and production of a node in the system.
    \item Many real-world trades have a unidirectional exchange relation. That is, the exchange might happen from $N_B$ to $N_A$, but not the other way around. Mutual arbitrage keeps the exchange value of the consuming node's credit stable in relation to the producing node's credit, as long as both nodes share other peers with one-another. If for example, $N_B$ only consumes from $N_A$ and not the other way around, but $N_C$ consumes from $N_B$, and $N_A$ consumes from $N_C$, then that circle of exchange relations will have an effect on the credit reserves of $E_{AB}$ after mutual arbitrage between $N_A$, $N_B$, and $N_C$. In other words, by trading to $N_C$, $N_B$ can increase its exchange value with $N_A$, even though their exchange relation is unidirectional.
\end{enumerate}

Let us denote with $R_{i,j}$ the credit reserve of $C_i$ within $E_{i,j}$. The way mutual arbitrage between three nodes $N_i$, $N_j$, and $N_k$ is performed, is by computing the optimal trade volume for a triangular arbitrage between the three nodes. In other words, each node wants to see if there exists a circular trade opportunity $C_1 \rightarrow C_2 \rightarrow C_3 \rightarrow C_1$ that results in more $C_1$ at the end of the triangular arbitrage, then was necessary at the beginning. To find the optimal trade volume for a triangular path (see Equation \ref{eq:optimal_arbitrage}.), each node calculates formula \ref{eq:optimal_arbitrage} for both the $C_1 \rightarrow C_2 \rightarrow C_3 \rightarrow C_1$ path, as well as for the $C_1 \rightarrow C_3 \rightarrow C_2 \rightarrow C_1$ path:

\begin{equation}
\label{eq:optimal_arbitrage}
 \Delta_{i} = \sqrt{a' a} - a
\end{equation}

Where $a'$ and $a$ can be defined as:

\begin{equation}
\label{eq:optimal_arbitrage_elaboration1}
        a' = \frac{R_{(i,k)} a'_{(k,i)}}{R_{(k,i)} + a'_{(k,i)}}; \qquad a = \frac{a'_{(i,k)} R_{(k,i)}}{R_{(k,i)} + a'_{(k,i)}};
\end{equation}

$a'_{(k,i)}$ and $a'_{(i,k)}$ can be defined as:

\begin{equation}
\label{eq:optimal_arbitrage_elaboration1}
        a'_{(k,i)} = \frac{R_{(j,i)} R_{(k,j)}}{R_{(j,k)} + R_{(j,i)}}; \qquad a'_{(i,k)} = \frac{R_{(i,j)} R_{(j,k)}}{R_{(j,k)} +  R_{(j,i)}};
\end{equation}

If one of the paths for $\Delta_i$ result in a value larger than zero, then $\Delta_i$ for the given path represents the optimal trade volume to initiate a triangular arbitrage. A mathematical proof for Equation \ref{eq:optimal_arbitrage} can be found in \cite{wang_cyclic_2022} work. Note that since Hypersyn's mutual arbitrage procedure is an agent-centric process, we can omit the fee variable used by constant product market makers like Uniswap. Once every node has computed the path that results in a positive $\Delta_i$, $\Delta_j$, $\Delta_k$, the smallest positive $\Delta$ is chosen as the value to initiate the triangular arbitrage between the three nodes.

If we take figure \ref{fig:mutual_arbitrage} as an example, the path $C_A \rightarrow C_B \rightarrow C_C \rightarrow C_A$ will result in the lowest positive $\Delta$. In this example, $N_A$ ends up inserting $\Delta_A$ amounts of $C_A$ into $E_{AB}$, removing a certain amount of $C_B$ from $E_{AB}$. The removed amount of $C_B$ is then used as input for the $E_{BC}$ edge, causing the removal of a certain amount of $C_C$ in the process. Finally, the removed amount of $C_C$ is then used as input for the $E_{AC}$ edge, removing a certain amount of $C_A$ and with it closing the arbitrage circle. At the end of this process, the edges between $N_A$, $N_B$, $N_C$ end up with new credit reserves, and therefore with new exchange values.

\begin{figure}[h]
\centering\includegraphics[width=0.3\linewidth]{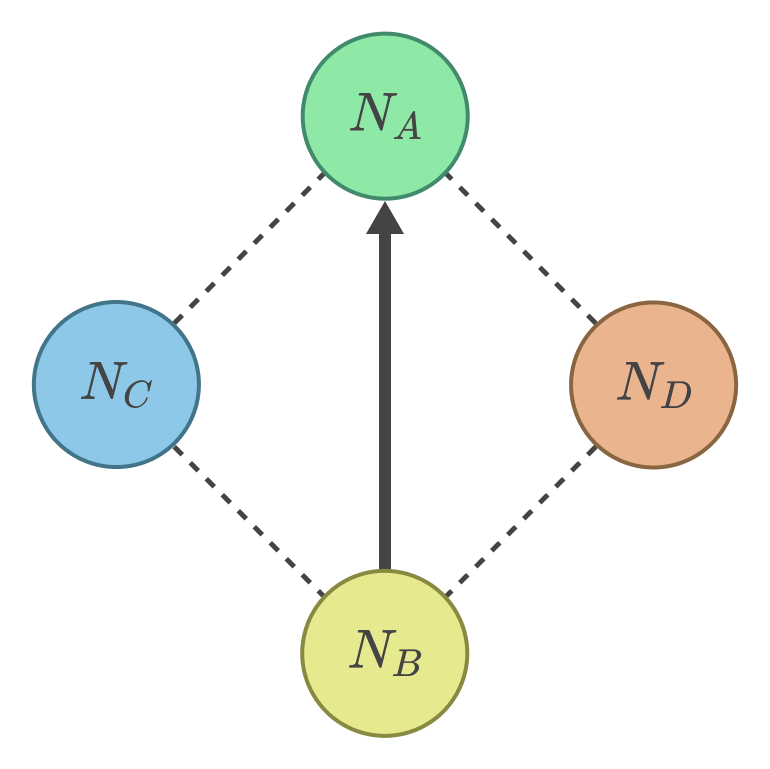}
\caption{A depiction of a node requesting multiple mutual arbitrages in Hypersyn. before $N_B$ transacts with $N_A$, it asks $N_A$ to adjust their credit reserves by performing mutual arbitrage with other nodes, once between ${N_A, N_B, N_C}$ and once between ${N_A, N_B, N_D}$.
}
\label{fig:multiple_mutual_arbitrage}
\end{figure}

The way mutual arbitrage is initiated has also important security implications. To prevent malicious price manipulations before a transaction takes place, the payment receiver decides on the node to participate in mutual arbitrage. If for example $N_B$ wants to increase its credit exchange value in relation to $N_A$'s credit (as in figure \ref{fig:multiple_mutual_arbitrage}), $N_B$ will ask $N_A$ to pick a common node (in this case $N_C$) for the mutual arbitrage. This means that $N_A$ will have to pick a peer whose optimal arbitrage has a "clockwise" cycle, e.g. $C_A \rightarrow C_B \rightarrow C_C \rightarrow C_A$, instead of $C_A \rightarrow C_C \rightarrow C_B \rightarrow C_A$, as choosing a peer with an "anti-clockwise" cycle would result in $N_B$'s credit exchange value to further decrease. If no common peer exists that $N_A$ trusts enough, which can participate in a favorable mutual arbitrage for $N_B$, the trade between $N_A$ and $N_B$ will be performed using the existing credit reserves of $E_{AB}$.

\subsection{State Model}
\label{ss: state model}
In Hypersyn, every node has a SMT as a state model, the leaves of which are populated with the cryptographic hashes of the node's edges. Each Hypersyn node, besides having a SMT, also has a counter $m$. Whenever the Merkle root of a node's SMT is updated, $m$ needs to be incremented.

Each edge consists of a reserve $R_i$ and $R_j$, and the two counters $m_i$ and $m_j$ that were used during the last state change of $R_i$ and $R_j$. The edge hash $H_{ij}$ is calculated as $H_{ij} = merkle(R_i,R_j, P_i,P_j, m_i,m_j)$, where $P_i$ and $P_j$ are the hashed public addresses of the nodes $N_i$ and $N_j$, and "merkle" simple returns the Merkle root for the array of variables. Whenever a trade or mutual arbitrage is performed, the edge state within the SMT is updated, which results in a new sparse Merkle root.

The path of an edge hash within the SMT is however not determined by $H_{ij}$ itself. The path within a tree is determined through the hash of the two nodes $U_{ij} = h(P_i,P_j)$. Or when compared to the concept of hash tables, $U_{ij}$ serves as a key within the SMT, whereas $H_{ij}$ as a value. This is done to prevent outdated edge states from being removed at one part of the tree, and new edge states being inserted in another part of the tree. By using $U_{ij}$ to determine the path of an edge within the tree, every edge update requires only a local update, instead of a deletion followed by an insertion.

Besides a counter, each edge also contains a signature field. The signature field of an edge consist of two tuples: $(S(i), p_i(j))$ and $(S(j), p_j(i))$, where $p_i(j)$ represents the Merkle proof of $SMT_i$ for $H_{ij}$, and $S$ represents:

\begin{equation}
\label{eq:edge_signature}
       S(i) = s_i(h(Root_i, m_i)) 
\end{equation}

where $s_i$ denotes the private signature of a Hypersyn node $N_i$, and $h$ represents a hash function. The reason for computing $s_i(h(Root_i, m_i))$ instead of $s_i(Root_i)$, is to have a potential proof of misbehavior. The reason why the information of the Merkle root is cryptographically signed, instead of the edge hash, is to increase the load of transactions that a Hypersyn node can process. Instead of signing every single update, a Hypersyn node can update several edges in parallel and sign its hashed Merkle root only once.

In Hypersyn, every node also stores a pruned version of each peer's SMT. If $L_i$ denotes the set of nodes to which $N_i$ is connected to, and $L_j$ represents the set of nodes to which $N_j$ is connected to, then the pruned copy of $SMT_j$ that $N_i$ will store, will be:

\begin{equation}
\label{eq:edge_signature}
       PSMT_j = L_j \cap L_i
\end{equation}

Where PSMT stands for pruned sparse Merkle tree. Note that the Merkle root of a pruned SMT will be the same as that of the original SMT. 

Whenever the sparse Merkle root of a PSMT is not the same as that of the sparse Merkle root of the peer, an anti-entropy process \cite{decandia_dynamo_2016} between the local PSMT and the peer's SMT is initiated. At the end of state synchronization, the root of the locally stores PSMT has to be the same as that of the peer's SMT root. This step has to be performed anytime before a transaction, or mutual arbitrage, if the roots are not the same.

Whenever the state of a locally stored PSMT is updated, any peer that is also connected to the node whose PSMT state changed, gets messaged and state synchronization gets performed. If after anti-entropy the roots of the two PSMTs are not the same, the leaf that is causing the inconsistency is found and every peer gets notified of the inconstancy, by providing the Merkle proof, counter, and signature for the given inconsistent edges. Every peer deletes their edges to the two nodes that caused the inconsistency as punishment for malicious behavior.

\subsection{Edge Formation \& Deletion}
\label{ss: edge formation}

Edges in the Hypersyn network can form in two ways:
\begin{enumerate}
    \item \textbf{Edge formation through a common node} - If $N_i$ and $N_j$ want to form a connection with one another, and both have a common connection with $N_k$, then we can use Equation \ref{eq:const_prod1} to determine the initial reserve ratios required for the new edge $E_{ij}$. That is, by looking at the reserve ratio of $E_{ik}$ and the reserve ratio of $E_{jk}$, we can determine what the reserve ratio of $E_{ij}$ has to be. This can be done by setting a temporary $E_{ij}$ with equal credit reserves and then calculating the mutual arbitrage between nodes $N_i$, $N_j$, and $N_k$.
    
    \item \textbf{Edge formation without any information} - In the case of nodes that share no common node with one-another, there is no way to know deterministically the ideal reserve ratio for the new edge. In such cases, the nodes are asked to negotiate an initial, acceptable exchange rate for their credit. Over time, the more edges a nodes forms, through mutual arbitrage, the exchange value of its credit to other peers changes accordingly to their rate of consumption and production within the system.
\end{enumerate}

Besides deleting edges due to malicious activity (as discussed in section \ref{ss: state model}), Hypersyn nodes can also delete the old edges that they share with peers that that have not performed mutual arbitrage for a certain amount of time. If for example the difference between the latest $m_j$ and $E_{ij}$'s $m_j$ counter is larger than a specific system-wide constant $v$, or in other words, $N_i$ was not able to reach and perform mutual arbitrage with $N_j$ for a long time, then $N_i$ can remove $H_{ij}$ from its SMT.

If $N_i$ performs a malicious edge deletion towards $N_j$ however, that is, $N_i$ deletes its edge with $N_j$ before their counter difference reaches $v$, and without having a proof of misbehavior for $N_j$, $N_j$, or any other node with a valid PSMT state, will contact $N_i$'s peers and forward them the proof. After this step, every node with the given proof of malicious deletion deletes its node with $N_i$.




\subsection{Hypersyn files}
\label{ss: network}

Hypersyn nodes use a gossip protocol to communicate with connected peers (i.e. with peers that have a common edge). For first time communications, Hypersyn makes use of a Kademlia \cite{goos_kademlia_2002} distributed hash table (DHT) variant that stores "Hypersyn files", which are similar in concept to torrent files. Every Hypersyn node has a Hypersyn file on the DHT, where the location of the file in the network is not determined by the file's hash, but by a hashed public key. This slight change allows us to have a mutable Hypersyn file. Each Hypersyn file contains the eventually consistent sparse Merkle root of the given Hypersyn node, its corresponding counter, and its corresponding signature (the information required for Equation \ref{eq:edge_signature}). Nodes in the DHT that store a particular Hypersyn file, update its content when a valid $S(i)$ is provided that has a larger $m_i$ than the previous version. In the original Kademlia design, files uploaded onto the DHT had to be refreshed by the original uploader after a specific time. In Hypersyn's Kademlia variant, any node can refresh the state of a Hypersyn file, as long as they can prove that they have an edge with $N_i$, i.e. they can provide a signed Merkle proof (signed by $N_i$) showing that they have an edge not older than $v$ with the given node. A Hypersyn file, besides containing the information required to compute $S(i)$, also contains a list of node IDs (the hash of a node's public hash) and IP addresses for each peer that has a connection to $N_i$. The given peer addresses can then be used as first contacts by new Hypersyn nodes, that want to form connections (or send proofs of misbehavior) to $N_i$'s peers. Before accepting a peer within a Hypersyn file as valid, a SMT presence proof is requested.

\section{Discussion}
\label{s: discussion}

This early version of the Hypersyn protocol is partially transparent in nature. That is, the peers which are connected to a Hypersyn node can know both the credit amount of the node's payments, as well as the credit reserves of the node with other with other shared peers. In upcoming protocol versions, we will explore the possibility of using zero knowledge proofs to hide this kind of data. As long as peers can ensure that 
\begin{enumerate}
    \item the product of the two reserves equals a system-wide constant $k$ without having knowledge of the reserves or input amount, and that
    \item the optimal trade volumes for mutual arbitrage can be properly compared in size to one another, without having to share any knowledge of the optimal trade volume,
\end{enumerate}
then the system could function in a completely private manner. 

Hypersyn, besides having the potential to substitute centrally-issued fiat/credit systems, also has the potential to be used as a layer 2 (L2) solution to facilitate frictionless transactions for blockchain-based systems. Smart contract exchanges could enable users to lock-in for example ether, in exchange for a Hypersyn mutual credit "ether". Nodes within the Hypersyn-Ethereum overlay network could trade Hypersyn ether without having to pay for gas (blockchain transaction fees). Each hosted token in this case would require a Hypersyn overlay network of its own. One can imagine further use-cases, such as the creation of peer-to-peer token exchanges between Hypersyn overlay networks.

Lastly, even though Hypersyn's goal is to make centrally-issued fiat systems obsolete, the network can nonetheless be of immense value for states. Central banks could setup a Hypersyn "super node", i.e. a node with a large number of peers. Instead of having to use a specific currency for trade between states, one state's super node could simply form a connection to another state's super node. Hypersyn has therefore the potential not only to increase the autonomy and self organization of people, but also enable the formation of a global peer-to-peer payment system for both states as well as productive entities. Such state super nodes would help in large-scale, fast, mutual arbitrage with its citizens, allowing for productive optimization. Furthermore, states without their own currency could tax their citizens in terms of e.g. labor time credit, rather than externally controlled currencies, effectively reducing foreign reserve pressure on the central bank and making taxes repayable in terms of a locally controlled denominator, e.g. Labor time.

\section{Conclusion}
\label{s: conclusion}

Hypersyn combines several well studied concepts to create the first agent-centric peer-to-peer system for mutual credit. The self-issued credit within the system does not have to be backed by a specific commodity, or a basket of commodities, and the credit exchange value of a peer gets dynamically adjusted by using an agent-centric approach inspired by the constant product market maker model. Hypersyn is permissionless, requires very little resources to run, and can easily process millions of transactions per second. More importantly, Hypersyn has all the necessary attributes for productive entities to exchange with one-another, without having to depend on the shortcomings of a centrally-issued fiat system.

\section{Acknowledgement}
I would like to thank Taulant Ramabaja for his constant support, invaluable feedback, and stimulating discussions. Without Taulant's guidance in the field of economics, the ideas necessary for the Hypersyn protocol would have never been formed. A debt of gratitude is also owed to Dr. Werner Retschitzegger from the Johannes Kepler University in Linz, who provided critical feedback on the structure of this work.

\section{Comments}
The Hypersyn protocol is being developed by 
\href{https://www.opncbr.com}{\underline{\textcolor{MidnightBlue}{Open Cybernetics}}}. For ideas and proposals to this work, feel free to make a \href{https://github.com/LumRamabaja/Hypersyn-paper}{\underline{\textcolor{MidnightBlue}{pull request}}} on the official GitHub repository of this paper.

\bibliographystyle{unsrt} 
\bibliography{references}

\end{document}